\documentclass[a4paper, 10pt, twocolumn]{article}

\usepackage[ansinew]{inputenc}
\usepackage[english]{babel}

\usepackage{graphicx}               
\usepackage{tabularx}               
\usepackage{multirow}               
\usepackage{url}                

\usepackage[small,bf]{caption2}     

\usepackage{parskip}

\usepackage{titlesec}
\usepackage{amsmath}                

\usepackage{commands}
\usepackage{import}
\usepackage{color}

\usepackage{microtype}
\usepackage[colorlinks = true,
linkcolor = black,
urlcolor  = black,
citecolor = black,
anchorcolor = black]{hyperref}

\titleformat{\section}{\normalfont\large\bfseries}{\thesection}{}{}
\titleformat{\subsection}{\normalfont\large\bfseries}{\thesection}{}{}
\titleformat{\paragraph}{\normalfont\bfseries}{\theparagraph}{}{}
\titlespacing{\section}{0pt}{6pt}{-1pt}
\titlespacing{\subsection}{0pt}{3pt}{-1pt}
\titlespacing{\paragraph}{0pt}{3pt}{-1pt}

\newcolumntype{Y}{>{\centering\arraybackslash}X}    

\let\svthefootnote\thefootnote
\newcommand\freefootnote[1]{%
	\let\thefootnote\relax%
	\footnotetext{#1}%
	\let\thefootnote\svthefootnote%
}

\begin{document}
\setlength{\abovedisplayskip}{5pt}
\setlength{\belowdisplayskip}{5pt}

\date{}                                         

\title{\vspace{-8mm}\textbf{\large
The PerspectiveLiberator – An Upmixing 6DoF Rendering Plugin\\[-4pt]
for Single-Perspective Ambisonic Room Impulse Responses}}

\author{
  Kaspar Müller, Franz Zotter\\
  \emph{\small%
    Institute of Electronic Music and Acoustics, University of Music and Performing Arts Graz, Austria
  }\\[-1pt]
  \emph{\small%
    Email: kaspar.mueller@outlook.de, zotter@iem.at
  }%
}\maketitle

\thispagestyle{empty}

\vspace*{-9mm}%
\section*{Introduction}
\label{sec:intro}
\vspace*{2pt}
Virtual reality interfaces allow the user to interactively change the position and look direction in six degrees of freedom (6DoF), and consistently with the visual part, the acoustic perspective needs to be updated. For audio rendering based on directional room impulse responses, the spatial impulse response rendering (SIRR) \cite{merimaa2004spatial} and (Ambisonic) spatial decomposition method (SDM)~\cite{tervo2013,zaunschirm2018brir}, as well as higher-order SIRR (HO-SIRR)~\cite{mccormack2020} were helpful to obtain rendering of sufficient resolution when using single-perspective room impulse responses. We can assume that these methods accomplish a reliable single-perspective rendering with dynamic head rotation.
Existing 6DoF audio rendering approaches using Ambisonic room impulse responses (ARIRs) or similar are, e.g., interpolation of B-format signals \cite{mariette2009,rivasmendez2018}, Vienna Multi Impulse Response (MIR)$^1$\!,
time warping \cite{masterson2009,garciagomez2018},
the 4-pi sampling reverberator~\cite{nakahara2019}, and higher-order parametric perspective extrapolation~\cite{mccormack2021}. 

\begin{figure}[htb]
  \centering
  \def\svgwidth{1.01\linewidth}\hspace*{-0.8mm}%
  \small\sffamily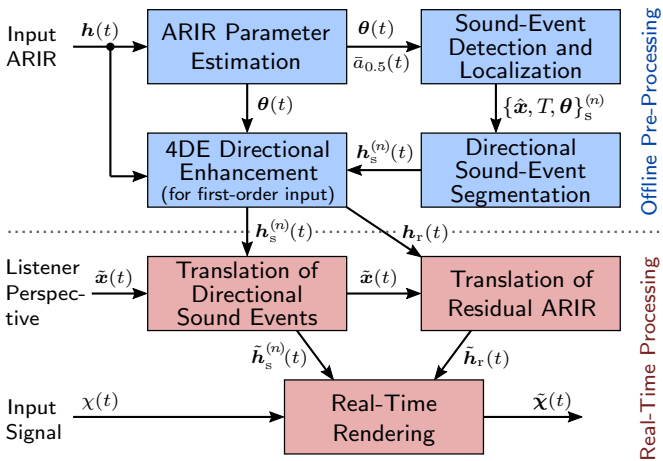
  \vspace*{-5mm}%
  \normalfont
  \caption{Signal flow graph of the PerspectiveLiberator.}
  \label{fig:signal_flow_graph}
\end{figure}

Based on the ARIR extrapolation in \cite{mueller2020_journal}, our contribution presents a plug-in meant to free the virtual perspective from the measured one: \emph{The PerspectiveLiberator}.
Figure~\ref{fig:signal_flow_graph} shows the two processing stages:\\
\emph{Offline pre-processing} detects and localizes sound events corresponding to the most distinct early ARIR peaks, and it removes according segments from the residual ARIR. The spatial resolution of such peak segments is increased by 4DE upmixing \cite{hoffbauer2020} to higher order Ambisonics (HOA).\\
\emph{Real-time processing} extrapolates the ARIR by time-, level- and directional-aligned translation of the sound-event segments according to the desired listener perspective and delays the residual ARIR to arrive after the direct sound peak (first sound event).
Finally, the HOA output signal is obtained by convolution-based rendering.

\newpage
\vspace*{-11mm}
\begin{figure}[hbt]
  \begin{center}
    \includegraphics[width=0.98\linewidth]{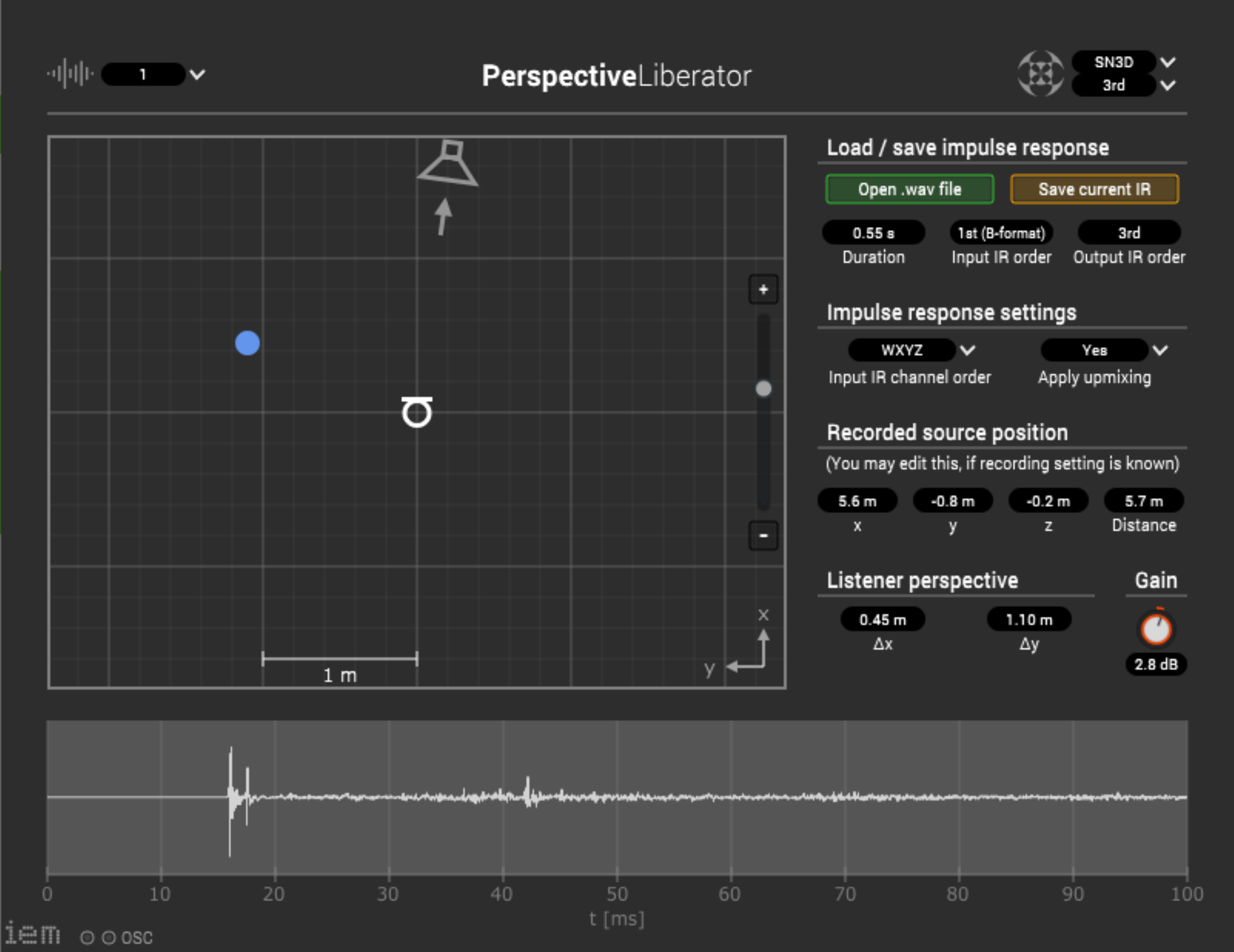}
  \end{center}
  \vspace*{-2.5mm}%
  \caption{User interface of the PerspectiveLiberator plug-in.}
  \vspace*{-5mm}%
  \label{fig:dode}
\end{figure}

\vspace{4mm}%
\section*{First-Order ARIR Parameter Estimation}
\label{sec:parameter_estimation}
\vspace*{3pt}%

\freefootnote{\!\!\!$^1$\url{https://www.vsl.co.at/de/Vienna_Software_Package/Vienna_MIR_PRO}}
As shown in \fig{signal_flow_graph}, the estimation of ARIR parameters is the initial step of the pre-processing stage.
Since most recorded ARIRs are first-order, we perform the parameter estimation based on first-order ARIR data only, even though any higher-order input is possible.
The two basic analysis features needed for the subsequent processing are the direction of arrival and the short-time amplitude.

\vspace*{1mm}%
\paragraph*{Direction of Arrival}
\label{sec:doa}
\vspace*{2pt}%

We compute a Cartesian direction of arrival (DOA) vector for each first-order input ARIR sample from the time averaged pseudo intensity vector (PIV) after band limitation
\vspace*{-12pt}%
\begin{flalign}
  \hat{\m{\theta}}(t) &{\,=\,} 
  \frac{\tilde{\m{\hspace*{-2pt}I}}(t) }
  {\Vert \hspace*{2pt}\tilde{\m{\hspace*{-2pt}I}}(t) \Vert } \,, \; \text{with} \;
  \,\tilde{\m{\hspace*{-2pt}I}}(t) {\,=\,}  F\lt{av}
  \Biggl\{ \hspace*{-1pt} W\lt{\scriptscriptstyle BP}(t)
  \M{
    \scriptstyle X\lt{BP}(t)\\[-2.5pt]
    \scriptstyle Y\lt{BP}(t)\\[-2.5pt]
    \scriptstyle Z\lt{BP}(t)
  }\hspace*{-1pt}\Biggr\} ,
  \label{eq:doa}
\end{flalign}
where $W\lt{\scriptscriptstyle BP}(t)$ is the omnidirectional, and $\brac{X,Y,Z}\lt{\scriptscriptstyle BP}\hspace*{-2pt}(t)$ are the first-order directional signals of $\m{h}(t)$, zero-phase bandpass filtered between 200\,Hz and 3\,kHz.
$F\lt{av}\brac{\cdot}$ is a zero-phase averaging filter over $0.25\,$ms.

\vspace*{1mm}%
\paragraph*{Short-Time Amplitude}
\label{sec:short_time_amplitude}
\vspace*{2pt}%

To obtain a meaningful ARIR amplitude estimate, which is especially needed for a robust peak detection, we propose to incorporate the directional ARIR content represented by the broadband PIV magnitude, instead of solely relating to the omnidirectional amplitude signal.
Time averaging of the PIV with a Hamming-windowed moving-average filter over 0.5\,ms, denoted by $F\lt{H} \!\brac{\cdot}$, yields
\vspace*{-2pt}%
\begin{align}
  \bar{a}_{0.5}(t) = 
  \sqrt{ F\lt{H} \big\{ \Vert \m{I}(t) \Vert \big\} } ,\;\;
  \text{with }\, \m{I}(t) = W(t) \M{
    \scriptstyle X(t) \\[-2pt] 
    \scriptstyle Y(t) \\[-2pt]
    \scriptstyle Z(t)}\hspace*{-2pt}.
  \label{eq:short_time_amplitude}
\end{align}%

\section*{Sound Events}
\label{sec:sound_events}
\vspace*{2pt}%

To auralize a variable listener perspective, our main focus is to detect and translate ARIR segments that contain the most distinct sound events, i.e.~peaks in the early ARIR caused by the direct sound and early reflections.
For each sound event, we localize its source position and extract a sound-event segment containing its directional ARIR content by steering a broad-band modal MVDR beamformer to its sound-event DOA.

\vspace*{1.5mm}%
\paragraph*{Sound-Event Detection and Localization}
\label{sec:sound_event_detection}
\vspace*{2pt}%

We propose to detect prominent sound events based on the short-time magnitude of the first-order ARIR content (see.~Eq.\,\eqref{eq:short_time_amplitude}).
To this end, we define a total number of sound events to detect ($P{\,\leq\,}10$ was found to be sufficient for medium reverberant rooms; more is possible for large rooms with strong early reflections) and look for the $P$ greatest peaks of $\bar{a}_{0.5}(t)$ within the first 100\,ms after the direct sound peak, which is referenced by $n{\,=\,}1$.
To prevent misinterpreting possible post-oscillations of high-level peaks as separate peak, we enhanced the peak selection by further criteria such as e.g.~a minimum peak distance and by fitting the short-time amplitude with exponential decay threshold curves.
The time of arrival (TOA) $T\lt{s}\nn$ of the $n^\mathrm{th}$ detected sound event can be directly utilized to estimate a sound-event DOA from the time-varying DOAs of \eq{doa}
\begin{flalign}
  \m\theta\lt{s}\nn &= \hat{\m\theta}(T\lt{s}\nn)
  \label{eq:sound_event_doa}
  \vspace*{-2pt}
\end{flalign}
and furthermore to localize the sound-event position
\begin{flalign}
  \hat{\m{x}}\lt{s}\nn &= \m{x} + c \cdot T\lt{s}\nn \cdot \m\theta\lt{s}\nn
  \,\stackrel{\m{x}=\m{0}}{=}\,
  c \cdot T\lt{s}\nn \cdot \m\theta\lt{s}\nn,
\end{flalign}
where $c$ is the speed of sound.
For simplification, we always set the recording perspective $\m{x}$ to the origin.

\vspace*{1.8mm}
\paragraph{Extraction of Sound-Event ARIR Segments}
\label{sec:sound_event_segmentation}
\vspace*{2pt}%

To separate the detected sound events from the input ARIR, we propose to extract windows around sound events using an MVDR beamformer.
Simply cutting out whole ARIR sound-event segments (as only reasonably done for the direct sound segment) leaves gaps in the residual ARIR which can induce undesired sound coloration. By contrast, the MVDR approach ensures a gap-less residual ARIR where only the most distinct directional information is removed and diffuse or ambient sound is preserved in the residual.
Below, the segmentation is described for Ambisonic order $N{\,=\,}1$, but yet any higher order can be processed similarly. The $n^\text{th}$ ARIR segment around time instant $T\nn\lt{s}$ is obtained by a window function $w\nn(t)$
which tapers the segment by cosine-shaped slopes of 0.5\,ms length
\begin{flalign}
	\m{h}\nn(t)=w\nn(t-T\nn\lt{s})\;\m{h}(t-T\nn\lt{s})\,.
  \label{eq:segment_window}
\end{flalign}
The directional signal in the segment from $\m{\theta}\lt{s}\nn$ is found~as
\begin{flalign}	
  s\lt{d}\nn(t) &= \pi \, \mt{y}_{1}\hspace*{-2pt}(\m{\theta}\lt{s}\nn)\,\m{h}\nn(t)\,,
  \label{eq:directional_signal}
  \\
  \text{with }\, \m{y}_{\!N}(\m\theta) &= \big[
  Y_0^0, \; Y_1^{-1}\!, \; Y_1^0, \; Y_1^1, \; \cdots , \, Y_N^N
  \big]^{\!\mathrm{T}}\! (\m\theta), \nonumber
\end{flalign}
using real-valued spherical harmonics $Y_l^m(\m{\theta})$ and a first-order N3D-normalized ARIR $\m{h}(t)$ in ACN sequence~\cite{nachbar2011}.
Re-encoding of this directional signal into zeroth and first order spherical harmonics at the respective DOA
helps to define a first-order residual ARIR segment, i.e.\ the ARIR segment with the directional signal removed
\vspace*{-1pt}%
\begin{flalign}
  \m{h}\lt{r}\nn(t) &=
  \m{h}\nn(t)-\m{y}_{1}(\m\theta\lt{s}\nn)\; s\lt{d}\nn(t)\,.
  \label{eq:residual_segment}
\end{flalign}

\vspace*{-5pt}%
For segmentation, we define the window $w\nn(t)$ of each segment (see~Eq.\,\eqref{eq:segment_window}) to start about 0.5\,ms before $T\lt{s}\nn$.
To yield preferably long segments, we limit each segment window end by the subsequent relevant peak TOA in the directional signal $s\lt{d}\nn(t)$ respectively to maximum 5\,ms length.
Note that especially for the direct-sound segment, too short segment windows should be avoided.

\vspace*{1.5mm}
\paragraph{Robustness to Beamforming Misalignment}
\vspace*{2pt}%
While each ARIR segment represents a well-localized sound event, the extraction of a single directional signal might fail if the estimated sound-event DOA in Equation~\eqref{eq:sound_event_doa} is imprecise, e.g., if there are multiple DOAs. Then beamforming would be misaligned and content of the directional sound-event signal would leak into the residual ARIR.
We therefore enhanced the approach in Equations~\eqref{eq:directional_signal} and \eqref{eq:residual_segment} by steering MVDR beamformers to four directions in tetrahedral layout similar as in \cite{hoffbauer2020}, which are neither considered to be fully directional nor fully residual.
The matrix containing the direction vectors of a tetrahedron in prototype orientation
\begin{flalign}
	\m{\Theta}_{0} = \M{
		1 & \scriptstyle-\frac{1}{3}      & 
		\scriptstyle-\frac{1}{3}          & \scriptstyle-\frac{1}{3}\\
		0 & \scriptstyle \sqrt\frac{2}{3}  & 
		0 & \scriptstyle -\sqrt\frac{2}{3}\\
		0 & \scriptstyle\sqrt\frac{2}{9}  &
		\scriptstyle-\sqrt\frac{8}{9}     & \scriptstyle\sqrt\frac{2}{9}
	}
\label{eq:tetrahedron}
\end{flalign}
is rotated towards the estimated sound-event DOA \eqref{eq:sound_event_doa} by
\begin{flalign}
	\begin{gathered}
		\hspace*{-5pt}
		\m{\Theta}_s\nn \,{=}\, \m{R}_{z}(\az_s\nn) \, \m{R}_y({\textstyle\frac{\pi}{2}}{\,-\,}\ze_s\nn) \, \m{\Theta}_{0}
		=
		\big[
		\m\theta_1, \m\theta_2, \m\theta_3, \m\theta_4
		\big],\\
		\text{where } \m\theta_1 = \m\theta_s\nn
	\end{gathered}
	\label{eq:rotated_tetrahedron}
\end{flalign}
for each sound event segment.
The according rotation matrices are determined by the azimuth $\az_s\nn$ and zenith $\ze_s\nn$ of the Cartesian sound-event DOA vector ${\m{\theta}}_s\nn\,{=}\,[x,y,z]^\text{T}$
\begin{flalign}
	\hspace*{-4pt}%
	\m{R}_{z}(\az)\,{=}\hspace*{-1pt}\M{
		\scriptstyle\cos{\az}  & \hspace*{-5pt}\scriptstyle-\sin{\az}\hspace*{-5pt} & \scriptstyle0 \\[-2pt]
		\scriptstyle\sin{\az}  & \hspace*{-5pt}\scriptstyle\cos{\az}\hspace*{-5pt}  & \scriptstyle0 \\[-2pt]
		\scriptstyle0          & \hspace*{-5pt}\scriptstyle0\hspace*{-5pt}             & \scriptstyle1
	}\hspace*{-2pt}\,&,\;
	\m{R}_y(\ze)\,{=}\hspace*{-1pt}\M{
		\scriptstyle\cos{\ze}  & \hspace*{-4pt}\scriptstyle0\hspace*{-4pt}& \scriptstyle\sin{\ze} \\[-2pt]
		\scriptstyle 0         & \hspace*{-4pt}\scriptstyle1\hspace*{-4pt} & \scriptstyle 0           \\[-2pt]
		\scriptstyle-\sin{\ze} & \hspace*{-4pt}\scriptstyle0\hspace*{-4pt} & \scriptstyle\cos{\ze}
	}\!,
	\label{eq:rotation_matrices}
	\\
	\az = \arctan \textstyle\frac{y}{x} \,&,\;
	\ze = \arccos z \,.
	\nonumber
\end{flalign}
The directional signal vector of the $n^\text{th}$ segment containing the four $\m\Theta_s\nn$-aligned directional signals replaces the single directional signal in Equation~\eqref{eq:directional_signal} by
\begin{align}
	\begin{aligned}
		\m{s}\lt{d}\nn(t) &= 
		\pi \, \mt{Y}_{1}\hspace*{-2pt}(\m{\Theta}\lt{s}\nn)\,\m{h}\nn(t) 
		\label{eq:directional_signals}\\
		\text{where } \m{Y}_{\hspace*{-3pt}N}(\m\Theta_s\nn) &= \hspace*{-1pt}
		\M{
			\m{y}_{\!N}(\m\theta_1),\,\dots,\,
			\m{y}_{\!N}(\m\theta_4)
		}.
	\end{aligned}
\end{align}

\vspace*{-6pt}%
Ideally, the first element of $\m{s}\lt{d}\nn(t)$, which is steered towards the estimated sound-event DOA, contains the directional sound-event signal while the remaining three elements mostly contain diffuse or ambient signals.

\vspace*{-3pt}%
To identify cases of misaligned beamforming, the 10\,ms-median of the short-time amplitude (see~Eq.\,\eqref{eq:short_time_amplitude}) around each sound-event peak is compared to the magnitude of the residual ARIR segment in Equation~\eqref{eq:residual_segment} at the peak location. This ratio yields an exclusion factor 
\begin{flalign}
  \!\!\m{r}\nn {=} \scriptstyle \begin{bmatrix}
    \scriptstyle 1\\[-2pt]
    \scriptstyle \text{max}\{1-\tilde r\nn\!,\,0\}\\[-2pt]
    \scriptstyle \text{max}\{1-\tilde r\nn\!,\,0\}\\[-2pt]
    \scriptstyle \text{max}\{1-\tilde r\nn\!,\,0\}
    \vspace*{0pt}
  \end{bmatrix},\;\,
  \tilde r\nn &{=} \frac{
      \text{med}_{10} \big\{ \bar{a}_{0.5} \{\m{h}(t)\} \big\}\Big|_{t{=}T\nn\lt{s}}\!\!\!}{
      \bar{a}_{0.5}\{\m{h}\nn\lt{r}(0)\}}
\end{flalign}
to partially keep the 3 off-DOA signals as a part of the directional segment whenever the magnitude of a plain residual $\m{h}\nn\lt{r}(t)$ as in Equation~\eqref{eq:residual_segment} would be too large.

We accordingly modify the first-order directional sound-event segment by softly excluding its 3 off-DOA signals
\begin{flalign}
    \m{s}\lt{d}\nn(t) &= 
    \pi \; \mathrm{diag}\{\m{r}\nn\!\} \,   \mt{Y}_{1}\hspace*{-2pt}(\m{\Theta}\lt{s}\nn)\,\m{h}\nn(t) \,.
    \label{eq:directional_segment_signals}
\end{flalign}

\vspace*{1mm}
\paragraph{Directional Enhancement}
\vspace*{2pt}%

4DE upmixing of each directional sound-event segment to HOA can be achieved according to \cite{hoffbauer2020} by
\begin{flalign}
	\m{h}\ltn{s}{N}\nn(t)
	&= \m{Y}_{\hspace*{-3pt}N}(\m{\Theta}\lt{s}\nn) \; \m{s}\lt{d}\nn(t) , \label{eq:directional_segment_upmixed}
\end{flalign}
where $N$ denotes the desired Ambisonic order.

The first-order residual ARIR is obtained by extracting the first-order sound event segments from the input ARIR
\begin{align}
	\m{h}\ltn{r}{1}(t) &= \m{h}(t) - \textstyle\sum\nolimits_n  \m{h}\ltn{s}{1}\nn(t+T\nn\lt{s}).
\end{align}
Finally, higher-order re-encoding yields
\begin{flalign}
		\m{h}\ltn{r}{N}(t) =
		\pi\,\m{Y}_{\hspace*{-3pt}N}(\m{\Theta}_t)\,
		\mt{Y}_{\hspace*{-3pt}1} \!(\m{\Theta}_t)\, \m{h}\ltn{r}{1}(t),
	\label{eq:residual_upmixed}
\end{flalign}
where $\m\Theta_t$ are the tetrahedral directions rotated towards the continuously time-varying DOA (see Eq.~\eqref{eq:tetrahedron}\,-\,\eqref{eq:rotated_tetrahedron} with $\m\theta_1\,{=}\,\hat{\m\theta}(t)$).
Keeping these directions constant in segmented parts of $\m{h}\ltn{s}{1}\nn(t)$ was omitted for simplicity.
Below, all subscripts $N$ denoting the Ambisonic order of upmixed sound-event and residual segments are omitted for brevity.

To avoid spectral whitening of late reverberation in higher orders, where fast DOA fluctuations cause amplitude modulation,
we apply a spectral correction of the temporal envelope to the residual ARIR similar as in \cite{zaunschirm2018brir}.
In particular, we use the spectral energies of the original first-order ARIR $\m{h}(t)$ averaged over its four components to correct both the higher order spectra and the diffuse-field sensitivities of the zeroth and first orders, which most often require equalization \cite{hoffbauer2020}.

\vspace*{1mm}
\section*{ARIR Translation}
\label{sec:arir_translation}
\vspace*{3pt}%

\fig{signal_flow_graph} shows ARIR extrapolation depending on the interactive, variable listener perspective $\tilde{\m{x}}(t)$.
Unlike the ARIR analysis, segmentation and upmixing, its translation needs to be processed in real time.
\vspace*{-6pt}
\begin{figure}[h]
  \centering
  \def\svgwidth{0.8\linewidth}\hspace*{8mm}%
  \small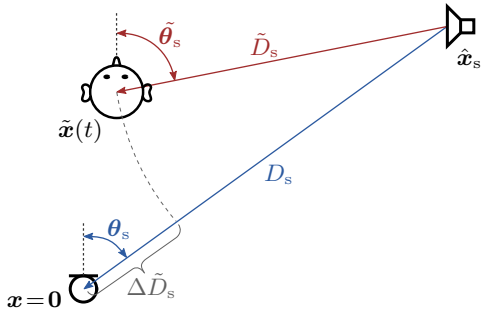
  \vspace*{-2pt}
  \caption{Translation of the listener $\m{x} \rightarrow \tilde{\m{x}}(t)$ in relation to a localized sound event position $\hat{\m{x}}\lt{s}$.}
  \label{fig:extrapolation}
\end{figure}

\vspace*{-1.5mm}%
\paragraph*{Real-Time Translation of Sound-Event Segments}
\label{sec:sound_event_translation}
\vspace*{2pt}%

This section describes the interactive translation of each sound-event segment by real-time manipulation of its 1+3 directional signals $\m{s}\lt{d}\nn(t)$
from Equation~\eqref{eq:directional_segment_upmixed}, according to the extrapolated listener perspective $\tilde{\m{x}}(t)$.

As displayed in \fig{extrapolation}, the extrapolation of the $n^\text{th}$ sound-event segment yields a rotation towards the corresponding sound-event position $\hat{\m{x}}\lt{s}\nn$ that is determined by the rotation matrix
\begin{flalign}
  \tilde{\m{R}}\nn &{\,=\,}
  \m{R}_z(\tilde\az\lt{s}\nn) \; \m{R}_y(\tilde\ze\lt{s}\nn{-}\ze\lt{s}\nn) \; \m{R}_z(-\az\lt{s}\nn)
  \label{eq:rotation}
\end{flalign}
with Equation \eqref{eq:rotation_matrices}.
Furthermore, the distance shift $\Delta \tilde{D}\lt{s}$ (see~Fig.\,\ref{fig:extrapolation}) is accompanying with a a gain \eqref{eq:distance_gain} due to the $\frac{1}{D}$ distance law, and a time shift \eqref{eq:time_shift} 
\begin{flalign}
	\tilde{g}\nn &= \frac{D\lt{s}\nn}{\tilde{D}\lt{s}\nn} =
	\frac{\Vert \hat{\m{x}}\lt{s}\nn \Vert}{
		\Vert \hat{\m{x}}\lt{s}\nn {-} \tilde{\m{x}}\nn\Vert},
	\label{eq:distance_gain}
	\\
	\Delta \tilde{t}\nn &= \textstyle \frac{1}{c} \big(D\lt{s}\nn {-} \tilde{D}\lt{s}\nn \big) =
	\frac{1}{c}
	\big( \Vert \hat{\m{x}}\lt{s}\nn \Vert \,{-}\, 
	\Vert \hat{\m{x}}\lt{s}\nn {-} \tilde{\m{x}}\nn\Vert \big),
	\label{eq:time_shift}  
\end{flalign}
where $c$ is the speed of sound.
To avoid critical translation positions, it is reasonable to limit the distance gain $\tilde{g}\nn$ to a maximum with a suitable soft-knee function. To moreover ensure that early reflection sound-event segments (indices of $n\,{\geq}\,2$) never occur earlier than the direct sound segment ($n\,{=}\,1$), time shifts in Equation~\eqref{eq:time_shift} are also limited to $T_s\nn{+}\Delta \tilde{t}^{\scriptscriptstyle(n)}{\,>\,}T_s^{\scriptscriptstyle(1)} {+} \Delta \tilde{t}^{\scriptscriptstyle(1)}$ using a lower-limited soft-knee function, such as e.g.~an arctan function.
Each translated ARIR sound-event segment is finally re-encoded in order $N$ from the directional signals $\m{s}\lt{d}\nn(t)$  of Equation~\eqref{eq:directional_segment_upmixed}, using its individual, position-dependent 
gain, rotation, and time delay
\begin{flalign}
  \tilde{\m{h}}\lt{s}\nn\!(t) = 
  \tilde{g}\nn \;
  \m{Y}_{\hspace*{-3pt}N}(\tilde{\m{R}}\nn\m{\Theta}\lt{s}\nn) \; \m{s}\lt{d}\nn\big(t\,{+}\,\Delta \tilde{t}\nn\big) .
  \label{eq:sound_event_translation}
\end{flalign}

\vspace*{2mm}%
\paragraph*{Real-Time Adaptation of the Residual ARIR}
\label{sec:residual_arir_translation}
\vspace*{2pt}%

The residual ARIR is shifted by the variable time shift of the direct-sound-event to ensure it begins thereafter
\begin{flalign}
  \tilde{\m{h}}\lt{r}(t) = \m{h}\lt{r}\big(t\,{+}\,\Delta \tilde{t}^{\scriptscriptstyle(\hspace*{-0.8pt}1\hspace*{-0.8pt})}\big).
\end{flalign}
As the most energetic directional segments, that are mainly responsible for the perceived localization, are excluded from our residual ARIR, we refrain from further and more elaborate processing of the residual to keep real-time computational costs at a minimum and furthermore to prevent the occurrence of possible additional sound-coloration artifacts.

The complete translated ARIR is the sum of the translated residual and the translated and time-shifted sound-events
\begin{flalign}
  \tilde{\m{h}}(t) = \tilde{\m{h}}\lt{r}(t) + \textstyle \sum\nolimits_n \tilde{\m{h}}\lt{s}\nn(t + T\lt{s}\nn) .
  \label{eq:translated_arir}
\end{flalign}

\vspace*{2mm}%
\paragraph*{Limits of the Movement Space}
\label{sec:movement_limits}
\vspace*{2pt}%

When virtually leaving the recorded room, jumps in the localization and sound coloration might appear due to large gains of possibly close localized early reflection sound events.
Therefore, it is meaningful to limiting the virtual movement space by virtual room walls.
These walls can be determined for instance based on the image-source model by the symmetry plane between the direct-sound source position $\hat{\m{x}}\lt{s}^{\scriptscriptstyle (\hspace*{-0.8pt}1\hspace*{-0.8pt})}$ and each reflection sound-event position $\hat{\m{x}}\lt{s}\nn\!,\, n\,{\geq}\,2$.
A listener perspective $\tilde{\m{x}}(t)$ within the virtual room thus corresponds with the condition
\begin{flalign}
  \big[\,\tilde{\m{x}}(t) - \big(\hat{\m{x}}\lt{s}\nn \!+ \m{v}\nn \big)\big]^\mathrm{T} \, {\m{v}\nn} \geq 0 \,, \; \forall\, n {\,\geq\,} 2 ,
\end{flalign}
where $\m{v}\nn {\,=\,} \frac{1}{2} (\hat{\m{x}}\lt{s}^{\scriptscriptstyle (\hspace*{-0.8pt}1\hspace*{-0.8pt})} {-\,} \hat{\m{x}}\lt{s}\nn )$ is the normal vector of a virtual room wall pointing inwards.

\section*{Efficient Real-Time 6DoF Rendering}
\label{sec:rendering}
\vspace*{3pt}%

\begin{figure}[hbt]
  \centering
  \def\svgwidth{0.95\linewidth}
  \small\sffamily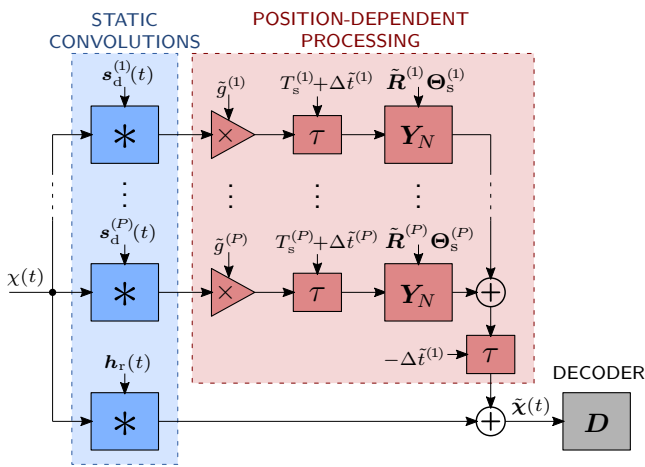%
  \vspace*{-2mm}%
  \normalfont
  \caption{\!Real-time\,6DoF\,rendering\,using\,static\,convolutions.}
  \label{fig:convolution}
\end{figure}

Instead of rendering the variable-perspective Ambisonic output signal $\tilde{\m\chi}(t)$ using a dynamic convolution of the single-channel input signal $\chi(t)$ with the variable ARIR $\tilde{\m{h}}(t)$ of Equation~\eqref{eq:translated_arir},
we suggest a rendering as shown in Figure~\ref{fig:convolution}.
Here, the convolution is split into separate static convolutions with the position-invariant directional sound-event signals (Eq.\,\eqref{eq:directional_segment_signals}), followed by the position-dependent translatory processing and upmixing of Equation~\eqref{eq:sound_event_translation}, and a convolution of the static upmixed residual (Eq.\,\eqref{eq:residual_upmixed}).
These convolutions can be processed using static multi-channel convolvers, which significantly reduces computational costs and implementation complexity.
Note, that in \fig{convolution} the output is additionally compensated for the variable delay $\Delta \tilde t^{\scriptscriptstyle (\hspace*{-0.8pt}1\hspace*{-0.8pt})}$, to reduce time-shift artifacts and to preserve a constant, position-independent residual ARIR.

\vspace*{-3mm}%
\section*{Conclusion}
\label{sec:conclusion}
\vspace*{3pt}%

In this paper, we described the fundamental approach of the \textit{PerspectiveLiberator} VST plug-in to achieve interactive 6DoF rendering based on real-time translated Ambisonic room impulse responses (ARIRs).\\
In the initial offline pre-processing stage, we estimated relevant first-order parameters in the input ARIR and extracted sound-event segments around the most prominent ARIR peaks containing distinct directional and temporal information using an MVDR beamformer.
Furthermore, we applied a directional enhancement to HOA.
Subsequently, we described the real-time variable-perspective translation of the sound-event segments according to their localized source positions and an appropriate adaption of the residual.
Finally, we described an efficient real-time rendering approach.

As proof of concept, we exported sound samples of linear listener movements through four different rooms\footnote{available at \url{https://phaidra.kug.ac.at/view/o:121102}}, in different layouts (rendered by sound-events / residual only as well as combined) with speech and guitar source~signals.

The perspective liberator plugin uses patented technology. Testing licenses will be issued by our company partner \textit{atmoky GmbH} upon request to \url{hello@atmoky.com}.

\setlength{\parskip}{7.2pt}
\bibliographystyle{daga}

\begin{thebibliography}{10}

\vspace*{5pt}%
\bibitem{merimaa2004spatial}
Merimaa, J. and Pulkki, V.: Spatial impulse response rendering.
\newblock Processings of the 7th International Conference on Digital Audio Effects, Naples, Italy, 2004.

\bibitem{tervo2013}
Tervo, S., P\"atynen, J., Kuusinen, A. and Lokki,~T.: Spatial decomposition
  method for room impulse responses.
\newblock Journal of the Audio Engineering Society 61.1 (2013), 17--28.

\bibitem{zaunschirm2018brir}
Zaunschirm, M., Frank, M. and Zotter, F.: {BRIR} synthesis using first-order
  microphone arrays.
\newblock 144th Audio Engineering Society Convention, 2018.

\bibitem{mccormack2020}
McCormack, L., Pulkki, V., Politis, A., Scheuregger, O. and Marschall, M.:
  Higher-order spatial impulse response rendering: Investigating the perceived
  effects of spherical order, dedicated diffuse rendering, and frequency
  resolution.
\newblock Journal of the Audio Engineering Society 68(5) (2020), 338--254.

\bibitem{mariette2009}
Mariette, N. and Katz, B.: {SoundDelta - Largescale, multi-user audio augmented
  reality}.
\newblock Proceedings of the EAA Symposium on Auralization (2009), 37--42.

\bibitem{rivasmendez2018}
M\'{e}ndez, D. R., Armstrong, C., Stubbs, J., Stiles, M. and Kearney, G.:
  Practical recording techniques for music production with six-degrees of
  freedom virtual reality.
\newblock 145th Audio Engineering Society Convention, 2018.

\bibitem{masterson2009}
Masterson, C., Kearney, G. and Boland, F.: Acoustic impulse response
  interpolation for multichannel systems using dynamic time warping.
\newblock Audio Engineering Society Conference.: 35th International Conference:
  Audio for Games, 2009.

\bibitem{garciagomez2018}
Garcia-Gomez, V. and Lopez, J. J.: Binaural room impulse responses interpolation
  for multimedia real-time applications.
\newblock 144th Audio Engineering Society Convention, 2018.

\bibitem{nakahara2019}
Nakahara, M., Omoto, A. and Nagatomo, Y.: Development of a 4-pi sampling
  reverberator, {VSVerb}. - {A}pplication to in-game sounds.
\newblock 146th Audio Engineering Society International Convention: Excite Your
  Ears, 2019.

\bibitem{mccormack2021}
McCormack, L., Politis, A. and Pulkki, V.: Parametric spatial audio effects
  based on the multi-directional decomposition of ambisonic sound scenes.
\newblock Proceedings of the 23rd International Conference on Digital Audio Effects (DAFx2020), Vienna, Austria, 2021.

\bibitem{mueller2020_journal}
M\"uller, K. and Zotter, F.: Auralization based on multi-perspective ambisonic
  room impulse responses.
\newblock Acta Acustica 4(6) (2020), 25.

\bibitem{hoffbauer2020}
Hoffbauer, E.: Development and evaluation of an algorithm for the enhancement
  of first-order ambisonic impulse responses.
\newblock Project thesis, 2020.

\bibitem{nachbar2011}
Nachbar, C., Zotter, F., Deleflie, E. and Sontacchi, A.: {AMBIX} - a suggested
  ambisonics format.
\newblock Ambisonics Symbosium, Lexington (KY), 2011.

\end{thebibliography}

\end{document}